\documentclass[aps,twocolumn,superscriptaddress,prd]{revtex4-2}

\usepackage{bm,graphicx,textcomp,amssymb,amsmath,dcolumn,color}
\usepackage{tikz,pgfplots}

\newcommand{\nn}{\nonumber\\}

\newcommand{\bea}{\begin{eqnarray}} 
\newcommand{\eea}{\end{eqnarray}}

\begin{document}

\title{Space-time toy model for Hawking radiation}

\author{Ralf Sch\"utzhold}

\affiliation{Helmholtz-Zentrum Dresden-Rossendorf, 
	Bautzner Landstra{\ss}e 400, 01328 Dresden, Germany,}

\affiliation{Institut f\"ur Theoretische Physik, 
	Technische Universit\"at Dresden, 01062 Dresden, Germany,}

\author{William G.~Unruh}

\affiliation{Department of Physics and Astronomy, 
University of British Columbia, Vancouver V6T 1Z1, Canada}

\affiliation{Hagler Institute for Advanced Study, Institute for Quantum Science 
and Engineering, Texas A\&M University, College Station, Texas 77843-4242, USA} 

\date{\today}

\begin{abstract}
By gluing together two sections of flat space-time in different metric 
representations (with the Minkowski metric representing the region far away 
from the black hole and the Rindler metric modeling the vicinity of the 
horizon), we construct a simplified toy model for black-hole evaporation.
The simple structure of this toy model allows us to construct exact analytic 
solutions for the two-point functions in the various vacuum states 
(Israel-Harte-Hawking, Unruh and Boulware states) in an easy way and 
thus helps to understand and disentangle the different ingredients for 
Hawking radiation better. 
\end{abstract}

\maketitle

%%%%%%%%%%%%%%%%%%%%%%%%%%%%%%%%%%%%%%%%%%%%%%%%%%%%%%%%%%%%%%%%%%%%%%%%%%%%%%%
%%%%%%%%%%%%%%%%%%%%%%%%%%%%%%%%%%%%%%%%%%%%%%%%%%%%%%%%%%%%%%%%%%%%%%%%%%%%%%%
\section{Introduction}
%%%%%%%%%%%%%%%%%%%%%%%%%%%%%%%%%%%%%%%%%%%%%%%%%%%%%%%%%%%%%%%%%%%%%%%%%%%%%%%
%%%%%%%%%%%%%%%%%%%%%%%%%%%%%%%%%%%%%%%%%%%%%%%%%%%%%%%%%%%%%%%%%%%%%%%%%%%%%%%

Hawking's striking discovery 
\cite{Hawking:1975vcx,Hawking:1974rv}
predicting that black holes should 
evaporate by emitting thermal radiation with the temperature 
\bea
{\mathfrak T}_{\rm Hawking}=\frac{\hbar c^3}{8\pi k_{\rm B} G_{\rm N} M}
\eea
suggests deep links between 
gravity ($G_{\rm N}$),
relativity ($c$),
quantum theory ($\hbar$) and 
thermodynamics ($k_{\rm B}$).
It seems as if nature is giving us a hint regarding its underlying structure. 
In order to interpret this hint correctly, it is important to properly 
understand the origin and mechanism of Hawking radiation.
The curvature of space-time is sometimes identified as one of the reasons or 
even the reason for Hawking radiation.
This idea could be supported by calculations \cite{Davies:1976ei}
of the trace anomaly 
of the energy-momentum tensor where one observes that the space-time curvature 
can be interpreted as a source term in the energy-momentum balance law, 
see also Eq.~\eqref{source-term} below. 

However, space-time curvature alone is not sufficient for predicting particle 
creation phenomena such as Hawking radiation. 
As a counter-example, one may consider the electromagnetic field around a 
neutron star in equilibrium which is described by a regular static metric.
%
%together with the electromagnetic field within this background. 
%
Unless there are photons incident from the outside, the electromagnetic field 
will quickly settle down to the local ground state in the vicinity of the star, 
and thus there is no lasting pair creation such as Hawking radiation. 
This is consistent with the trace anomaly calculation mentioned above, 
because a static metric and thus static curvature does not generate a source 
term in the balance law for energy -- corresponding to the $\nu=0$ component in 
Eq.~\eqref{source-term} below -- but only a source term for the momentum balance 
(i.e., the $\nu=1$ component) which can be interpreted as a force density. 

Since the total energy is conserved in static or stationary space-times, 
lasting particle creation phenomena such as Hawking radiation are only possible 
if there is some place where the energy of the created particles comes from. 
For Hawking radiation, this is the horizon -- the constant flux of positive 
energy out to infinity due to Hawking radiation
(from the point of view of static observers far away)
is compensated by the flux of negative energy into the horizon. 
However, this is still not the full picture since the quantum state near the horizon 
can be locally indistinguishable from vacuum while Hawking radiation is observed 
at infinity.
Furthermore, the quantum energy inequalities 
(see, e.g., \cite{Fewster:2012yh,Ford:1994bj}) 
demand that the 
region with negative energy cannot be arbitrarily large (where the precise 
meaning of the term ``arbitrarily large'' depends on the explicit form of 
the inequality under consideration). 

The transition from the local vacuum near the horizon and the thermal radiation 
observed far away is related to the space-time curvature, especially the spatial 
dependence of the red-shift. 
As a simplified intuitive picture, the finite pressure of the thermal radiation 
observed far away and the vanishing pressure of the local vacuum state near the 
horizon require some finite force density in between, which is generated by the 
curvature via the trace anomaly, see the $\nu=1$ component of 
Eq.~\eqref{source-term} below. 

Hence, Hawking radiation is caused by a combination of space-time 
curvature and the horizon. 
Since disentangling these effects is rather complicated for the Schwarzschild 
metric, we consider a simpler toy model in the following. 
To this end, we consider gluing together two regions of flat space-time, 
one representing the vicinity of the horizon and the other one spatial infinity, 
such that the curvature is restricted to the boundary between the two regions. 
The fact that we have a piece-wise flat space-time simplifies the analysis and allows 
us to discuss the different vacuum states by means of exact analytic solutions. 

%\end{document}

%%%%%%%%%%%%%%%%%%%%%%%%%%%%%%%%%%%%%%%%%%%%%%%%%%%%%%%%%%%%%%%%%%%%%%%%%%%%%%%
%%%%%%%%%%%%%%%%%%%%%%%%%%%%%%%%%%%%%%%%%%%%%%%%%%%%%%%%%%%%%%%%%%%%%%%%%%%%%%%
\section{The Model}   
%%%%%%%%%%%%%%%%%%%%%%%%%%%%%%%%%%%%%%%%%%%%%%%%%%%%%%%%%%%%%%%%%%%%%%%%%%%%%%%
%%%%%%%%%%%%%%%%%%%%%%%%%%%%%%%%%%%%%%%%%%%%%%%%%%%%%%%%%%%%%%%%%%%%%%%%%%%%%%%

As motivated above, let us consider the following 
metric in 1+1 dimensions ($\hbar=c=1$)
\bea
\label{metric}
ds^2
=
%\exp\left\{2\kappa x\Theta(-x)\right\}
%\left(dt^2-dx^2\right)
%=
\left\{
\begin{array}{lll}
e^{2\kappa x}(dt^2-dx^2) & {\rm for} & x<0 
\\
dt^2-dx^2 & {\rm for} & x>0 
\end{array}
\right.
\,,
\eea
where $\kappa$ corresponds to the surface gravity. 
For positive $x$, this metric just describes flat space-time in 
terms of the usual Minkowski coordinates $t$ and $x$, while for 
for negative $x$, it is related to the Rindler metric 
$ds^2=\kappa^2\rho^2d\tau^2-d\rho^2$ via the 
identification $t\leftrightarrow\tau$ and 
$x\leftrightarrow\rho_*=\ln(\kappa\rho)/\kappa$,
i.e., the Regge-Wheeler tortoise coordinate. 

In order to determine which part of the space-time is covered by these 
coordinates, let us recall the relations between Rindler $\tau,\rho$ 
and Minkowski coordinates $T,X$ with $ds^2=dT^2-dX^2$ which read 
$T=\rho\sinh(\kappa\tau)$ and $X=\rho\cosh(\kappa\tau)$. 
Hence, the region of negative $x$ corresponds to that part of the Rindler 
wedge where $0<\rho<1/\kappa$.
Thus, both $x>0$ and $x<0$ correspond to flat space-time, though in different 
coordinates.  
At the boundary $x=0$, however, we have an infinite curvature 
(which will become important below).
Note that this boundary $x=0$ is not the horizon, which would be at 
$x\to-\infty$, i.e., $\rho=0$. 

World-lines with constant $x>0$ just describe static observers in Minkowski 
space-time whereas world-lines with constant $x<0$ correspond to accelerated 
observers.
On the other hand, inertial observers starting at $x<0$ would either run to 
$x\to-\infty$ (i.e., reach the horizon) in a finite proper time or cross the 
boundary at $x=0$. 
In that region $x>0$, the world-lines of inertial observers are just the 
usual straight lines.  

For the two-point functions discussed in the next section, it is advantageous 
to introduce the light-cone variables $u=t-x$ and $v=t+x$ such that 
\bea
\label{metric-uv}
ds^2
=
\left\{
\begin{array}{lll}
e^{\kappa(v-u)}\,du\,dv & {\rm for} & v<u 
\\
du\,dv & {\rm for} & v>u 
\end{array}
\right.
\,.
\eea
In the Rindler wedge, i.e., for $x<0$, they are related to the standard 
Minkowski light-cone variables via 
%we may also use the 
%alternative representation 
$U=T-X=-e^{-\kappa u}/\kappa$ and 
$V=T+X=e^{\kappa v}/\kappa$ which gives $ds^2=dU\,dV$.
Note that the coordinates $T,X$ and $U,V$ are regular across the horizon 
$\rho=0$ and can be extended beyond it.  

%%%%%%%%%%%%%%%%%%%%%%%%%%%%%%%%%%%%%%%%%%%%%%%%%%%%%%%%%%%%%%%%%%%%%%%%%%%%%%%
%%%%%%%%%%%%%%%%%%%%%%%%%%%%%%%%%%%%%%%%%%%%%%%%%%%%%%%%%%%%%%%%%%%%%%%%%%%%%%%
\section{Two-Point Functions}
%%%%%%%%%%%%%%%%%%%%%%%%%%%%%%%%%%%%%%%%%%%%%%%%%%%%%%%%%%%%%%%%%%%%%%%%%%%%%%%
%%%%%%%%%%%%%%%%%%%%%%%%%%%%%%%%%%%%%%%%%%%%%%%%%%%%%%%%%%%%%%%%%%%%%%%%%%%%%%%

For simplicity, let us first consider a massless scalar field 
\bea
\label{massless}
\Box\phi=0
\,.
\eea
Due to its conformal invariance in 1+1 dimensions, we get the usual 
decoupling into left- and right-moving modes 
\bea
\phi(t,x)
&=&
\phi_{\rm left}(t+x)+\phi_{\rm right}(t-x)
\nn 
&=&
\phi_{\rm left}(v)+\phi_{\rm right}(u)
\,.
\eea
After quantization, this decoupling into left- and right-moving modes 
does also apply to the quantum field, which facilitates the discussion 
of the different vacuum states. 

%%%%%%%%%%%%%%%%%%%%%%%%%%%%%%%%%%%%%%%%%%%%%%%%%%%%%%%%%%%%%%%%%%%%%%%%%%%%%%%
\subsection{Israel-Hartle-Hawking Vacuum}  
%%%%%%%%%%%%%%%%%%%%%%%%%%%%%%%%%%%%%%%%%%%%%%%%%%%%%%%%%%%%%%%%%%%%%%%%%%%%%%%

Casting aside the problems related to the infrared divergence of the massless 
scalar field in 1+1 dimensions for a moment, the Minkowski vacuum in the 
Rindler wedge (for $x<0$) has the standard two-point function 
(away from the light cone) 
\bea
\label{IHH-vacuum}
\langle\hat\phi(t,x<0)\hat\phi(t',x'<0)\rangle
=
-\frac{1}{4\pi}\,\ln\left[\kappa^2\Delta U\Delta V\right]
\nn
=
-\frac{1}{4\pi}\,\ln\left[\kappa^2(T-T')^2-\kappa^2(X-X')^2\right]
%+{\rm const}
\nn
=
-\frac{1}{4\pi}\,\ln\left[
e^{\kappa(x+x')}\cosh(\kappa[t-t'])
-e^{2\kappa x}
-e^{2\kappa x'}
\right]
%+{\rm const}
\,,
\eea
where $\Delta U=U-U'$ and $\Delta V=V-V'$. 
Now, since this two-point function is a solution of the wave equation 
$(\partial_t^2-\partial_x^2)\phi=0$ for both arguments $t,x$ and $t',x'$ 
and all values of $x$ and $x'$, it has the same form 
-- in terms of the coordinates $t,x$ and $t',x'$ -- %(as in the second line)
on the other side $x>0$ where it describes a thermal state with the Unruh 
temperature ${\mathfrak T}_{\rm Unruh}=\kappa/(2\pi)$.
Hence, this state corresponds to the Israel-Hartle-Hawking state
\cite{Israel:1976ur,Hartle:1976tp}. 

%%%%%%%%%%%%%%%%%%%%%%%%%%%%%%%%%%%%%%%%%%%%%%%%%%%%%%%%%%%%%%%%%%%%%%%%%%%%%%%
\subsection{Boulware Vacuum} 
%%%%%%%%%%%%%%%%%%%%%%%%%%%%%%%%%%%%%%%%%%%%%%%%%%%%%%%%%%%%%%%%%%%%%%%%%%%%%%%

In contrast, let us start from the Minkowski vacuum in the other region $x>0$ 
\bea
\label{B-vacuum}
\langle\hat\phi(t,x>0)\hat\phi(t',x'>0)\rangle
=
-\frac{1}{4\pi}\,\ln\left[\kappa^2\Delta u\Delta v\right]
\nn
=
-\frac{1}{4\pi}\,\ln\left[\kappa^2(t-t')^2-\kappa^2(x-x')^2\right]
%+{\rm const}
\,.
\eea
With the same argument as before, this form remains correct in the 
Rindler wedge, i.e., for $x<0$, where it corresponds to the Rindler vacuum.
This state is the ground state of the Hamiltonian generating the time $t$ 
evolution, which is usually referred to as the Boulware vacuum
\cite{Boulware:1974dm}. 
In contrast to the Israel-Hartle-Hawking state above, this state becomes 
singular in terms of the $T$ and $X$ coordinates when approaching the 
horizon at $x\to-\infty$.  

%%%%%%%%%%%%%%%%%%%%%%%%%%%%%%%%%%%%%%%%%%%%%%%%%%%%%%%%%%%%%%%%%%%%%%%%%%%%%%%
\subsection{Unruh Vacuum}  
%%%%%%%%%%%%%%%%%%%%%%%%%%%%%%%%%%%%%%%%%%%%%%%%%%%%%%%%%%%%%%%%%%%%%%%%%%%%%%%

For a state corresponding to black-hole evaporation \cite{Unruh:1976db}, 
we take the left-moving 
modes to start in the Minkowski vacuum at $x>0$, given by the $v$-term in 
Eq.~\eqref{B-vacuum}, while the right-moving modes start in the Minkowski 
vacuum at $x<0$, given by the $u$-contribution in Eq.~\eqref{IHH-vacuum} 
\bea
\label{U-vacuum}
\langle\hat\phi(t,x)\hat\phi(t',x')\rangle
=
-\frac{1}{4\pi}
\left(
\ln\left[\kappa\Delta U\right]+\ln\left[\kappa\Delta v\right]
\right)
\nn
=
-\frac{1}{4\pi}
\left(
\ln\left[e^{-\kappa u'}-e^{-\kappa u}\right]+\ln\left[\kappa\Delta v\right]
\right)
\,.
\eea
Since the right-moving modes start in the Minkowski vacuum at $x<0$,
they are regular at the horizon. 
After propagating to the other side $x>0$, inertial observers perceive them 
as thermal radiation. 
The same inertial observers at $x>0$ would assign zero occupation numbers 
to the left-moving modes. 
However, after the left-moving modes propagate to the other side $x<0$,
they would no longer appear unoccupied for inertial observers. 
In fact, similar to the Rindler vacuum state, they would become singular 
at the (past) horizon. 

%%%%%%%%%%%%%%%%%%%%%%%%%%%%%%%%%%%%%%%%%%%%%%%%%%%%%%%%%%%%%%%%%%%%%%%%%%%%%%%
%%%%%%%%%%%%%%%%%%%%%%%%%%%%%%%%%%%%%%%%%%%%%%%%%%%%%%%%%%%%%%%%%%%%%%%%%%%%%%%
\section{Energy-Momentum Tensor}  
%%%%%%%%%%%%%%%%%%%%%%%%%%%%%%%%%%%%%%%%%%%%%%%%%%%%%%%%%%%%%%%%%%%%%%%%%%%%%%%
%%%%%%%%%%%%%%%%%%%%%%%%%%%%%%%%%%%%%%%%%%%%%%%%%%%%%%%%%%%%%%%%%%%%%%%%%%%%%%%

After having discussed the two-point functions, let us investigate the 
renormalized expectation value of the energy-momentum tensor 
$\langle\hat T^\mu_\nu\rangle_{\rm ren}$.
In principle, this quantity $\langle\hat T^\mu_\nu\rangle_{\rm ren}$ can be 
obtained via the point-slitting technique from the above two-point functions
(or other techniques). 
Here, we employ energy balance law 
$\nabla_\mu\langle\hat T^\mu_\nu\rangle_{\rm ren}=0$
which can be cast into the form 
\bea
\label{toy-model-balance}
%\nabla_\mu 
%\langle\hat T^\mu_\nu\rangle_{\rm ren}
%=
\frac{1}{\sqrt{-g}}\,\partial_\mu
\left(\sqrt{-g}\,\langle\hat T^\mu_\nu\rangle_{\rm ren} %T^\mu_\nu
\right)
=
\frac12\,
\langle\hat 
T^{\alpha\beta}
\rangle_{\rm ren}\,
\partial_\nu g_{\alpha\beta}
%=0
\,.
\eea
For the metric~\eqref{metric}, the right-hand side is proportional to the 
trace $\langle\hat T^\mu_\mu\rangle_{\rm ren}$. 
For the classical field~\eqref{massless}, this trace would be zero, 
but quantum fields in curved space-times acquire a trace anomaly
\cite{Davies:1976ei}, which is, 
for the massless scalar field, given by 
$\langle\hat T^\mu_\mu\rangle_{\rm ren}=R/(24\pi)$ in terms of the 
Ricci scalar $R$.  

For $x\neq0$, the metric~\eqref{metric} just corresponds to flat space-time 
and thus does not have any curvature, but at $x=0$, we get a delta singularity 
in $R$ and thus $\langle\hat T^\mu_\mu\rangle_{\rm ren}$ which can be 
interpreted as a source term in Eq.~\eqref{toy-model-balance}. 
Note that special care is required for computing this source term as the delta 
function in $\langle\hat T^\mu_\mu\rangle_{\rm ren}$ is multiplied with the 
metric derivative $\partial_\nu g_{\alpha\beta}$ which has a Heaviside like 
step at $x=0$.  
One way would be to start with a smooth metric and then take the appropriate 
limit, another option would be to split off the trace term and consider the 
trace-free part $\theta_{\mu\nu}$, see Sec.~\ref{Trace-free Tensor} below. 

However, symmetry arguments already allow us to draw some general 
conclusions at this stage:
Since the metric~\eqref{metric} is static and 
$\langle\hat T^\mu_\nu\rangle_{\rm ren}$ is stationary for the three 
vacuum states discussed above, all time-derivatives $\partial_t$ vanish 
in Eq.~\eqref{toy-model-balance}. 
Then, evaluating Eq.~\eqref{toy-model-balance} for $\nu=0$, we find that 
the normalized energy flux $\sqrt{-g}\,\langle\hat T^1_0\rangle_{\rm ren}$
(or momentum density) must always be constant across the whole space-time.
For $\nu=1$, Eq.~\eqref{toy-model-balance} shows that the effective pressure 
$\sqrt{-g}\,\langle\hat T^1_1\rangle_{\rm ren}$ is constant for $x>0$ and for 
$x<0$ but displays a step at $x=0$.  
As the trace $\langle\hat T^\mu_\mu\rangle_{\rm ren}$ vanishes for $x\neq0$,
%the same is true for 
the energy density $\sqrt{-g}\,\langle\hat T^0_0\rangle_{\rm ren}$
behaves in the same way, i.e., it is constant for $x>0$ and for 
$x<0$ but displays a step at $x=0$.  

%%%%%%%%%%%%%%%%%%%%%%%%%%%%%%%%%%%%%%%%%%%%%%%%%%%%%%%%%%%%%%%%%%%%%%%%%%%%%%%
\subsection{Israel-Hartle-Hawking Vacuum}  
%%%%%%%%%%%%%%%%%%%%%%%%%%%%%%%%%%%%%%%%%%%%%%%%%%%%%%%%%%%%%%%%%%%%%%%%%%%%%%%

Since the Israel-Hartle-Hawking vacuum %this state 
just corresponds to the Minkowski vacuum for $x<0$, we 
have $\langle\hat T^\mu_\nu\rangle_{\rm ren}=0$ in this region $x<0$. 
With the argument above, this already implies that the energy flux 
vanishes everywhere $\langle\hat T^1_0\rangle_{\rm ren}=0$. 
This should be no surprise because for $x>0$ the Israel-Hartle-Hawking 
vacuum is indistinguishable from a thermal state with the Unruh temperature 
where the left- and right-moving fluxes cancel each other. 
In this thermal region $x>0$, we find the usual energy density of a thermal 
bath $\langle\hat T^0_0\rangle_{\rm ren}\propto\kappa^2$. 
Thus energy density and pressure 
%$\langle\hat T^1_1\rangle_{\rm ren}=-\langle\hat T^0_0\rangle_{\rm ren}$
are positive for $x>0$ and jump to zero at $x=0$ where the jump in pressure 
is counter-balanced by the curvature singularity.  

%%%%%%%%%%%%%%%%%%%%%%%%%%%%%%%%%%%%%%%%%%%%%%%%%%%%%%%%%%%%%%%%%%%%%%%%%%%%%%%
\subsection{Boulware Vacuum}  
%%%%%%%%%%%%%%%%%%%%%%%%%%%%%%%%%%%%%%%%%%%%%%%%%%%%%%%%%%%%%%%%%%%%%%%%%%%%%%%

When going from the Israel-Hartle-Hawking state to the Boulware vacuum,
we have to subtract the thermal bath of particles.  
For $x>0$, this just means that $\langle\hat T^\mu_\nu\rangle_{\rm ren}$
is reduced to zero -- as expected in the Minkowski vacuum. 
On the other side $x<0$, however, this subtraction implies that the 
energy density $\langle\hat T^0_0\rangle_{\rm ren}$ must become negative: 
Here world-lines with constant $x$ correspond to uniformly accelerated 
observers who would experience the Minkowski vacuum 
(i.e., the Israel-Hartle-Hawking state discussed above) 
as a thermal bath of particles. 
From the point of view of those Rindler observers, the energy density 
$\langle\hat T^0_0\rangle_{\rm ren}$ can be split up into two parts -- 
the energy density of the Rindler vacuum and the energy density from 
the thermal bath of particles.  
The latter is always positive and thus removing it yields a negative 
energy density for the Rindler vacuum since we started from 
$\langle\hat T^\mu_\nu\rangle_{\rm ren}=0$ in the Minkowski vacuum.  
Hence, the energy density and pressure make the same step as before, 
but are globally shifted down. 
As before, energy flux must vanish since 
$\langle\hat T^\mu_\nu\rangle_{\rm ren}$ is zero for $x>0$.  

%%%%%%%%%%%%%%%%%%%%%%%%%%%%%%%%%%%%%%%%%%%%%%%%%%%%%%%%%%%%%%%%%%%%%%%%%%%%%%%
\subsection{Unruh Vacuum}  
%%%%%%%%%%%%%%%%%%%%%%%%%%%%%%%%%%%%%%%%%%%%%%%%%%%%%%%%%%%%%%%%%%%%%%%%%%%%%%%

A non-vanishing flux is obtained in the Unruh state where the incoming 
(i.e., left-moving) modes are in their vacuum state for $x>0$ while the 
outgoing (i.e., right-moving) modes are thermally occupied for $x>0$. 
This results in a flux of positive energy moving out to $x\to\infty$. 
Since $\sqrt{-g}\,\langle\hat T^1_0\rangle_{\rm ren}$ must be the same 
for all $x$, we also find an energy flux for $x<0$. 
From the point of view of the Rindler observers (at constant and negative $x$), 
the left-moving modes are in the Rindler vacuum state 
(i.e., unoccupied by particles) while the right-moving modes are thermally 
occupied. 
These thermal particles compensate the negative energy density of the 
Rindler vacuum itself for the right-moving modes, but they are absent for the 
left-moving modes, which means that we have a flux of negative energy to 
$x\to-\infty$.  

%%%%%%%%%%%%%%%%%%%%%%%%%%%%%%%%%%%%%%%%%%%%%%%%%%%%%%%%%%%%%%%%%%%%%%%%%%%%%%%
\subsection{Trace-Free Tensor}\label{Trace-free Tensor}
%%%%%%%%%%%%%%%%%%%%%%%%%%%%%%%%%%%%%%%%%%%%%%%%%%%%%%%%%%%%%%%%%%%%%%%%%%%%%%%

As already mentioned above, an alternative way of interpreting the results is 
to define a trace-free tensor $\theta_{\mu\nu}$ by splitting off the trace 
anomaly \cite{Davies:1976ei}
\bea
\langle\hat T_{\mu\nu}\rangle_{\rm ren}
=
\theta_{\mu\nu}+\frac{R}{48\pi}\,g_{\mu\nu}
\,,
\eea
such that $\theta_\mu^\mu=0$. 
In terms of $\theta_{\mu\nu}$, the energy balance law 
$\nabla_\mu\langle\hat T^\mu_\nu\rangle_{\rm ren}=0$ becomes 
\bea
\label{source-term}
\nabla_\mu\theta_\nu^\mu
=
\frac{1}{\sqrt{-g}}\,\partial_\mu\left(\sqrt{-g}\,\theta^\mu_\nu\right)
=
-\frac{\partial_\nu R}{48\pi}
\,,
\eea
where the role of the curvature as the source term becomes even more apparent.  
For a metric of the form $ds^2=C(u,v)\,du\,dv$ such as in Eq.~\eqref{metric},
point-splitting renormalization yields the following result 
\bea
\theta_{uu}
=
-\frac{\sqrt{C}}{12\pi}\,\frac{\partial^2}{\partial u^2}\frac{1}{\sqrt{C}}
\,,\quad 
\theta_{vv}
=
-\frac{\sqrt{C}}{12\pi}\,\frac{\partial^2}{\partial v^2}\frac{1}{\sqrt{C}}
\,,\,
\nn
\theta_{uv}=\theta_{vu}=0
\,,
\eea
for the vacuum state corresponding to the $u$ and $v$ coordinates, 
which is the Boulware vacuum state in our case. 

Inserting the metric in Eqs.~\eqref{metric} and \eqref{metric-uv}, 
we find that $\theta_{\mu\nu}$ vanishes for $x>0$ as expected.
For $x<0$, on the other hand, we find the constant values 
$\theta_{uu}=\theta_{vv}=-\kappa^2/(48\pi)$. 
Apart from the jump at $x=0$, these terms do also contain delta singularities 
at $x=0$. 

For the Israel-Hartle-Hawking vacuum, this negative energy density in the 
$x<0$ region is shifted up to zero by adding a constant to the energy 
density $\sqrt{-g}\,\langle\hat T^0_0\rangle_{\rm ren}$ 
(in terms of the original $t$ and $x$ coordinates) 
such that we arrive at a positive energy density for $x>0$. 

For the Unruh vacuum, this shift is only applies to the right-moving modes, 
while the left-moving modes remain unaffected. 
As a result, we only obtain half the energy density for $x>0$ in comparison 
to the Israel-Hartle-Hawking vacuum.
Furthermore, this asymmetry between left- and right-moving modes induces a 
non-zero energy flux, as discussed above. 

%%%%%%%%%%%%%%%%%%%%%%%%%%%%%%%%%%%%%%%%%%%%%%%%%%%%%%%%%%%%%%%%%%%%%%%%%%%%%%%
%%%%%%%%%%%%%%%%%%%%%%%%%%%%%%%%%%%%%%%%%%%%%%%%%%%%%%%%%%%%%%%%%%%%%%%%%%%%%%%
\section{Potential Barrier}  
%%%%%%%%%%%%%%%%%%%%%%%%%%%%%%%%%%%%%%%%%%%%%%%%%%%%%%%%%%%%%%%%%%%%%%%%%%%%%%%
%%%%%%%%%%%%%%%%%%%%%%%%%%%%%%%%%%%%%%%%%%%%%%%%%%%%%%%%%%%%%%%%%%%%%%%%%%%%%%%

So far, we considered the case of a massless scalar field in 1+1 dimensions, 
which is conformally invariant. 
As a result, the left- and right-moving modes decouple -- which greatly 
simplifies the analysis. 
In the general case, however, the conformal invariance is broken, e.g., 
by a finite mass term, fields with higher spin, or the angular barrier 
in 3+1 dimensions. 
These effects typically induce back-scattering, i.e., the scattering from 
left- to right-moving modes and vice versa.   

In order to incorporate back-scattering into our toy model, we introduce 
an additional delta potential at $x=0$ 
\bea
\label{barrier}
\Box\phi=\gamma\delta(x)\phi 
\,.
\eea
In terms of the usual mode decomposition for the initial quantum fields 
\bea
\hat\phi^{\rm in}_{\rm left}
=
\int d\omega\,
\frac{e^{-i\omega(t+x)}}{\sqrt{4\pi\omega}}\,
\hat a_\omega^{\rm left}
+{\rm h.c.}
\,,\quad 
\nn
\hat\phi^{\rm in}_{\rm right}
=
\int d\omega\,
\frac{e^{-i\omega(t-x)}}{\sqrt{4\pi\omega}}\,
\hat a_\omega^{\rm right}
+{\rm h.c.}
\,,
\eea
the potential barrier then induces the reflection and transmission 
coefficients 
\bea
\hat a_\omega^{\rm left}
\to 
{\mathcal T}_\omega\hat a_\omega^{\rm left}+
{\mathcal R}_\omega\hat a_\omega^{\rm right}
%\nn
\,,\quad 
\hat a_\omega^{\rm right}
\to 
{\mathcal T}_\omega^*\hat a_\omega^{\rm right}+
{\mathcal R}_\omega^*\hat a_\omega^{\rm left}
\,,
\eea
with $1/{\mathcal T}_\omega=i\gamma/(2\omega)+1$ 
and $1/{\mathcal R}_\omega=2i\omega/\gamma-1$
for the simple delta potential~\eqref{barrier}, 
but other potentials can be treated in complete analogy. 

This allows us to calculate all expectation values for the final modes 
in terms of expectation values of the initial operators 
$\hat a_\omega^{\rm left}$ and $\hat a_\omega^{\rm right}$
such as 
\bea
\langle(\hat a_\omega^{\rm left})^\dagger\hat a_\omega^{\rm left}\rangle
\to
|{\mathcal T}_\omega^2|
\langle(\hat a_\omega^{\rm left})^\dagger\hat a_\omega^{\rm left}\rangle
+
|{\mathcal R}_\omega^2|
\langle(\hat a_\omega^{\rm right})^\dagger\hat a_\omega^{\rm right}\rangle
\nn
%&&
+\left[{\mathcal T}_\omega^*{\mathcal R}_\omega
\langle(\hat a_\omega^{\rm left})^\dagger\hat a_\omega^{\rm right}\rangle
+{\rm h.c.}
\right].
\quad 
\eea
In the cases considered here, the initial modes incident from left and right 
are uncorrelated such that the mixed terms in the square bracket on the 
right-hand side vanish.  
For the Boulware and the Israel-Hartle-Hawking states, the two expectation 
values 
$\langle(\hat a_\omega^{\rm left})^\dagger\hat a_\omega^{\rm left}\rangle$
and 
$\langle(\hat a_\omega^{\rm right})^\dagger\hat a_\omega^{\rm right}\rangle$ 
give the same result -- in the first case, both vanish and in the second 
case, both yield the same thermal distribution.   
Together with unitarity $|{\mathcal T}_\omega^2|+|{\mathcal R}_\omega^2|=1$,
we find that all local expectation values (or, more generally, all 
expectation values confined to one side) yield the same results as 
in the case without the potential.  
In contrast, correlations  between the two sides will be generated by the 
potential.  

In the Unruh vacuum, however, the two expectation values differ:
$\langle(\hat a_\omega^{\rm left})^\dagger\hat a_\omega^{\rm left}\rangle$
vanishes while 
$\langle(\hat a_\omega^{\rm right})^\dagger\hat a_\omega^{\rm right}\rangle$ 
yields a thermal distribution.   
Thus the positive energy flux going out to $x\to\infty$ will be reduced by 
the gray-body factor 
$|{\mathcal T}_\omega^2|
\langle(\hat a_\omega^{\rm right})^\dagger\hat a_\omega^{\rm right}\rangle$ 
while the flux of negative energy to $x\to-\infty$ is partially compensated 
by the reflected thermal radiation 
$|{\mathcal R}_\omega^2|
\langle(\hat a_\omega^{\rm right})^\dagger\hat a_\omega^{\rm right}\rangle$.  

As an intuitive picture, one can say that the negative flux of energy 
into the horizon is created by the absence of the particles which would be 
flowing into the black hole in the Israel-Hartle-Hawking state.
The escape of particles to $x\to\infty$ instead of their reflection back 
into the future horizon creates that negative flux. 

%%%%%%%%%%%%%%%%%%%%%%%%%%%%%%%%%%%%%%%%%%%%%%%%%%%%%%%%%%%%%%%%%%%%%%%%%%%%%%%
%%%%%%%%%%%%%%%%%%%%%%%%%%%%%%%%%%%%%%%%%%%%%%%%%%%%%%%%%%%%%%%%%%%%%%%%%%%%%%%
\section{Conclusions and Outlook}  
%%%%%%%%%%%%%%%%%%%%%%%%%%%%%%%%%%%%%%%%%%%%%%%%%%%%%%%%%%%%%%%%%%%%%%%%%%%%%%%
%%%%%%%%%%%%%%%%%%%%%%%%%%%%%%%%%%%%%%%%%%%%%%%%%%%%%%%%%%%%%%%%%%%%%%%%%%%%%%%

\begin{figure}[tbp]
\includegraphics[height=0.2\textheight]{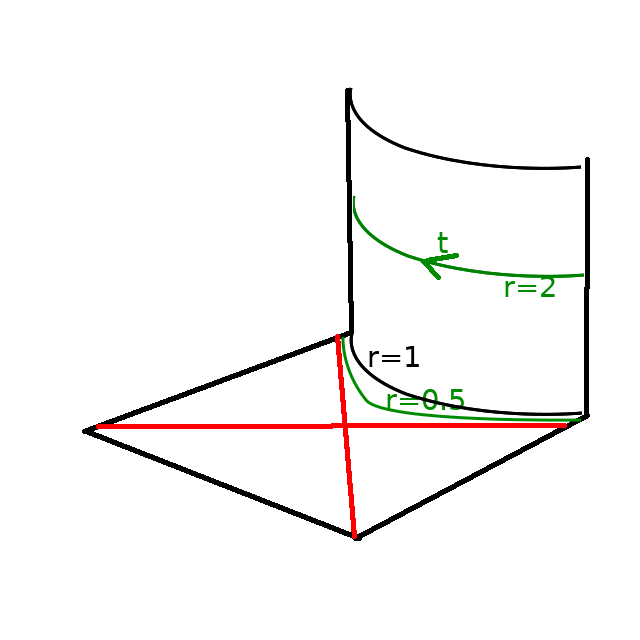}
\hspace{1cm}
\includegraphics[height=0.2\textheight]{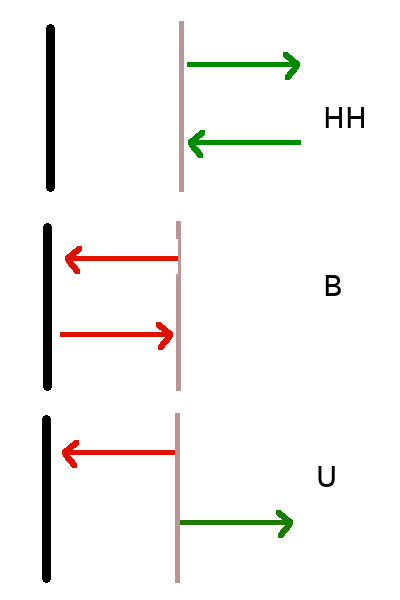}
\hspace{1cm}
\includegraphics[height=0.2\textheight]{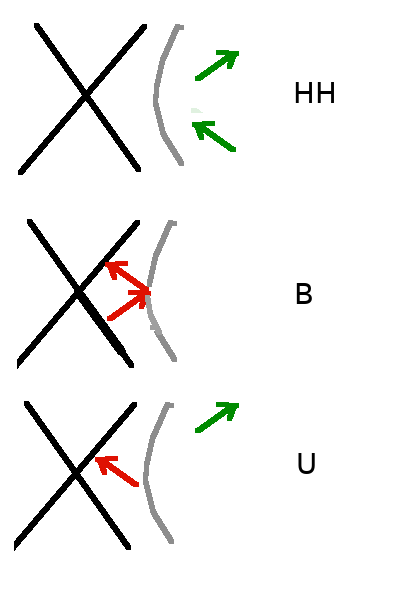}
\caption{Left: three-dimensional visualization of the two-dimensional
space-time~\eqref{metric} where the horizontal planar surface representing 
the near-horizon region $x<0$ and the 
(extrinsically curved but intrinsically flat) 
vertical surface representing the outside region $x>0$
are glued together at the effective radius $r=1$.  
Middle and right: sketch of the fluxes in the Israel-Hartle-Hawking (top), 
Boulware (middle) and Unruh (bottom) vacua.}
\label{fig:exmp}
\end{figure}

For a better understanding of complex phenomena in Nature, it is often useful 
to construct suitable toy models (see, e.g., \cite{Maia:2007vy}) which 
reproduce essential features of the original phenomenon. 
Here, we consider the phenomenon of Hawking radiation, i.e., black-hole 
evaporation. 
In order to disentangle the roles played by space-time curvature and 
horizon, we construct a toy model by gluing together two patches of 
piece-wise flat space-times in Rindler and Minkowski coordinates. 
This simplified space-time allows us to provide compact analytic solutions 
for the two-point functions and the energy-momentum tensor for the 
Israel-Hartle-Hawking, Unruh and Boulware vacua, see Fig.~\ref{fig:exmp}. 

In the Unruh and Boulware vacua, we find a negative energy density in the 
Rindler patch, which can be explained by the absence of Rindler particles 
which would lift the energy density up to zero in the 
Israel-Hartle-Hawking vacuum (which is just the Minkowski vacuum inside 
this Rindler patch). 
Note that, consistent with the energy inequalities, inertial observers 
cannot stay forever in the Rindler patch, they either fall into the 
black hole or move out to the Minkowski patch in a finite proper time.

In order to mimic the curvature potential of black holes, we also 
considered a potential barrier in the form of a simple delta potential. 
For Israel-Hartle-Hawking and Boulware vacua, this additional potential 
does not change the particle spectra 
(only the correlations between different sides of the delta potential change).  
For the Unruh vacuum, on the other hand, we find a gray-body factor 
for the outgoing radiation, as expected. 

Instead of the most simple example of a scalar field considered here, 
one could generalize our studies to massless fermions in 1+1 dimensions,
see, e.g., 
\cite{Davies:1977yw,DeWitt:2003pm,Christensen:1978yd,Ferrero:2023unz}. 
There are two main differences.  
First, the total flux is obtained by an integral over the 
Fermi-Dirac distribution 
(with $\exp\{\omega/(k_{\rm B}{\mathfrak T}_{\rm Hawking})\}+1$ 
in the denominator) 
instead of the Bose-Einstein (or Planck) distribution 
(with $\exp\{\omega/(k_{\rm B}{\mathfrak T}_{\rm Hawking})\}-1$ 
in the denominator), which yields a reduction by a factor of two.
Second, we have to sum over more species in the fermionic case. 
On the one hand, particles and anti-particles both contribute 
equally to the total 
flux\footnote{The same would happen for a complex scalar field.} 
and, on the other hand, we have to add up the spin species 
(depending on the realization, e.g., $2\times2$ or $4\times4$
Dirac matrices). 
Nevertheless, up to the resulting pre-factor, % (a suitable power of two), 
we obtain the same result for trace anomaly. 
Thus, the qualitative results for energy density and flux as well as 
pressure for the three vacuum states under consideration should be 
equivalent. 

%%%%%%%%%%%%%%%%%%%%%%%%%%%%%%%%%%%%%%%%%%%%%%%%%%%%%%%%%%%%%%%%%%%%%%%%%%%%%%%
%%%%%%%%%%%%%%%%%%%%%%%%%%%%%%%%%%%%%%%%%%%%%%%%%%%%%%%%%%%%%%%%%%%%%%%%%%%%%%%
\section*{Acknowledgements}  
%%%%%%%%%%%%%%%%%%%%%%%%%%%%%%%%%%%%%%%%%%%%%%%%%%%%%%%%%%%%%%%%%%%%%%%%%%%%%%%
%%%%%%%%%%%%%%%%%%%%%%%%%%%%%%%%%%%%%%%%%%%%%%%%%%%%%%%%%%%%%%%%%%%%%%%%%%%%%%%

%\bigskip 

I first met Renaud Parentani, %Reynaud Parentani, 
to whom this paper is dedicated, about 20 years ago at Peyresq. 
He was always someone who had thought deeply about anything that he talked 
about, and what he said was well worth listening to.
He was also someone who enjoyed food -- I learned saffron (and where the best
place in Valencia to buy it) and truffles from him. 
He was also someone who had a hard time listening to someone who had not 
thought through what they were saying, as could be seen by the scowl on 
his face and his restlessness. 
I will certainly miss him and his insights.
\hfill 
William G.~Unruh

\bigskip 

After having read several of his inspiring papers, I also met 
Renaud Parentani at various conferences and workshops.  
Apart from his passion for physics and other topics of interest to him 
(e.g., how to buy promising young wine and to let it mature), 
I was impressed by his command and precision in language -- 
in science and otherwise. 
I remember one anecdote where he was asked about his nationality and said: 
``I disguise as a french.'' 
The interlocutor (not a native speaker) did not quite understand and asked: 
``You mean `pretend'?'', to which Renaud Parentani smiled and replied: 
``Not quite the same.''
Not having him around anymore is a real loss. 
\hfill 
Ralf Sch\"utzhold 

\bigskip 

We acknowledge fruitful discussions with Sebastián Franchino-Viñas. 
R.S.~acknowledges funding by the Deutsche Forschungsgemeinschaft 
(DFG, German Research Foundation) -- Project-ID 278162697-- SFB 1242.

%$%$

%\newpage 


\begin{thebibliography}{99}

%\cite{Hawking:1975vcx}
\bibitem{Hawking:1975vcx}
S.~W.~Hawking,
``Particle Creation by Black Holes,''
Commun. Math. Phys. \textbf{43}, 199-220 (1975)
[erratum: Commun. Math. Phys. \textbf{46}, 206 (1976)]
doi:10.1007/BF02345020
%11485 citations counted in INSPIRE as of 05 Sep 2024

%\cite{Hawking:1974rv}
\bibitem{Hawking:1974rv}
S.~W.~Hawking,
``Black hole explosions,''
Nature \textbf{248}, 30-31 (1974)
doi:10.1038/248030a0
%4632 citations counted in INSPIRE as of 05 Sep 2024

%\cite{Davies:1976ei}
\bibitem{Davies:1976ei}
P.~C.~W.~Davies, S.~A.~Fulling and W.~G.~Unruh,
``Energy Momentum Tensor Near an Evaporating Black Hole,''
Phys. Rev. D \textbf{13}, 2720-2723 (1976)
doi:10.1103/PhysRevD.13.2720
%350 citations counted in INSPIRE as of 05 Sep 2024

%\cite{Fewster:2012yh}
\bibitem{Fewster:2012yh}
C.~J.~Fewster,
``Lectures on quantum energy inequalities,''
[arXiv:1208.5399 [gr-qc]].
%70 citations counted in INSPIRE as of 05 Sep 2024

%\cite{Ford:1994bj}
\bibitem{Ford:1994bj}
L.~H.~Ford and T.~A.~Roman,
``Averaged energy conditions and quantum inequalities,''
Phys. Rev. D \textbf{51}, 4277-4286 (1995)
doi:10.1103/PhysRevD.51.4277
%[arXiv:gr-qc/9410043 [gr-qc]].
%244 citations counted in INSPIRE as of 05 Sep 2024

%\cite{Israel:1976ur}
\bibitem{Israel:1976ur}
W.~Israel,
``Thermo field dynamics of black holes,''
Phys. Lett. A \textbf{57}, 107-110 (1976)
doi:10.1016/0375-9601(76)90178-X
%614 citations counted in INSPIRE as of 05 Sep 2024

%\cite{Hartle:1976tp}
\bibitem{Hartle:1976tp}
J.~B.~Hartle and S.~W.~Hawking,
``Path Integral Derivation of Black Hole Radiance,''
Phys. Rev. D \textbf{13}, 2188-2203 (1976)
doi:10.1103/PhysRevD.13.2188
%1120 citations counted in INSPIRE as of 05 Sep 2024

%\cite{Boulware:1974dm}
\bibitem{Boulware:1974dm}
D.~G.~Boulware,
``Quantum Field Theory in Schwarzschild and Rindler Spaces,''
Phys. Rev. D \textbf{11}, 1404 (1975)
doi:10.1103/PhysRevD.11.1404
%391 citations counted in INSPIRE as of 05 Sep 2024

%\cite{Unruh:1976db}
\bibitem{Unruh:1976db}
W.~G.~Unruh,
``Notes on black hole evaporation,''
Phys. Rev. D \textbf{14}, 870 (1976)
doi:10.1103/PhysRevD.14.870
%4186 citations counted in INSPIRE as of 05 Sep 2024

%\cite{Maia:2007vy}
\bibitem{Maia:2007vy}
C.~Maia and R.~Sch\"utzhold,
``Quantum toy model for black-hole back-reaction,''
Phys. Rev. D \textbf{76}, 101502 (2007)
doi:10.1103/PhysRevD.76.101502
%[arXiv:0706.4010 [gr-qc]].
%13 citations counted in INSPIRE as of 05 Sep 2024

%\cite{Davies:1977yw}
\bibitem{Davies:1977yw}
P.~C.~W.~Davies and W.~G.~Unruh,
``Neutrino Stress Tensor Regularization in Two-Dimensional Space-Time,''
Proc. Roy. Soc. Lond. A \textbf{356}, 259-268 (1977)
doi:10.1098/rspa.1977.0131
%14 citations counted in INSPIRE as of 05 Sep 2024

%\cite{DeWitt:2003pm}
\bibitem{DeWitt:2003pm}
B.~S.~DeWitt,
``The global approach to quantum field theory. Vol. 1, 2,''
Int. Ser. Monogr. Phys. \textbf{114}, 1-1042 (2003)
%98 citations counted in INSPIRE as of 05 Sep 2024

%\cite{Christensen:1978yd}
\bibitem{Christensen:1978yd}
S.~M.~Christensen,
``Regularization, Renormalization, and Covariant Geodesic Point Separation,''
Phys. Rev. D \textbf{17}, 946-963 (1978)
doi:10.1103/PhysRevD.17.946
%354 citations counted in INSPIRE as of 05 Sep 2024

%\cite{Ferrero:2023unz}
\bibitem{Ferrero:2023unz}
R.~Ferrero, S.~A.~Franchino-Vi\~nas, M.~B.~Fr\"ob and W.~C.~C.~Lima,
``Universal Definition of the Nonconformal Trace Anomaly,''
Phys. Rev. Lett. \textbf{132}, no.7, 071601 (2024)
doi:10.1103/PhysRevLett.132.071601
%[arXiv:2312.07666 [hep-th]].
%4 citations counted in INSPIRE as of 05 Sep 2024

\end{thebibliography}
\end{document}